\documentclass[aps,prl,reprint,groupedaddress,amsmath,amssymb,floatfix,graphics, graphicx,subfigure]{revtex4-2}
\usepackage{todonotes}
\usepackage{color}

\usepackage[utf8]{inputenc}

\begin{document}

\title{\bf Coordinated Stress-Structure Self-Organization in Granular Packing}

   \author{Xiaoyu Jiang}
   \affiliation{Department of Engineering Mechanics and Energy, University of Tsukuba, Tsukuba, Ibaraki, 305-8573, Japan}
 \author{Raphael Blumenfeld}\email{rbb11@cam.ac.uk}
    \affiliation{Gonville \& Caius College, University of Cambridge, UK}
   \author{Takashi Matsushima}\email{tmatsu@kz.tsukuba.ac.jp}
   \affiliation{Department of Engineering Mechanics and Energy, University of Tsukuba, Tsukuba, Ibaraki, 305-8573, Japan}

\date{\today}

\begin{abstract}
During quasi-static dynamics of granular systems, the stress and structure self-organise, but there is currently no quantitative measure or understanding of this phenomenon.
Such an understanding is essential because local structural properties of the settled material are then correlated with the local stress, which calls into question existing linear theories of stress transmission in granular media.  
A method to quantify the local stress-structure correlations is necessary for addressing this issue and we present here such a method for planar systems. We then use it to analyze numerically several different systems, compressed quasi-statically by two different procedures. 
We define cells, cell orders, cell orientations, and cell stresses and report the following results. 
1. Cells orient along the local stress major principal axes. 
2. The mean ratio of cell principal stresses decreases with cell order and increases with friction.
3. The ratio distributions collapse onto a single curve under a simple scaling, for all packing protocols and friction coefficients.
4. A constructed model explains the correlations between the local cell and stress principal axis orientations. 
5. The collapse of the stress ratios onto a Weibull distribution is explained theoretically.
Our results quantify the cooperative stress-structure self-organization and provide a way to relate quantitatively the stress-structure coupling to different process parameters and particle characteristics.
Significantly, the strong stress-structure correlation, driven by structural re-organization upon application of external stress, suggests that current stress theories of granular matter need to be revisited.
\end{abstract}

\maketitle

\begin{figure}[!h]
    \includegraphics[width=0.8\linewidth]{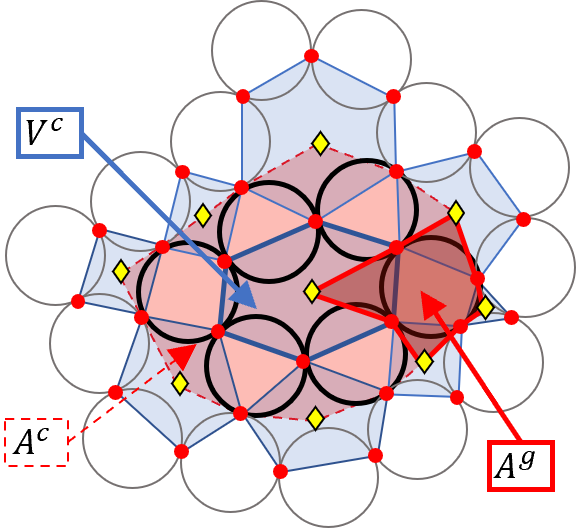}
    \caption{Cell definition and structure. 
    Central blue polygon: the cell \cite{matsushima2014universal,matsushima2017fundamental,BaBl2002} of area $V^c$, surrounded by the discs shown in thick black lines; 
    red points: contacts; 
    yellow diamonds: cell centers, defined as the mean positions of the contacts surrounding a cell;
    dark red polygon: the area associated with a disc, $A^g$, used to calculate the disc stress; 
    dashed red polygon: the sum over the disc areas $A^g$ around the cell, $A^c$, used to calculate the cell stress.}
    \label{fig:cellStructure}
\end{figure}

Granular matter plays an important role in nature and in our everyday life, but understanding its rich and complex behavior is far from complete. One difficulty is that the sensitivity of macroscopic behavior to grain-scale structural characteristics hinders conventional coarse-graining methods. A fundamental difficulty is that theories for mechanics of such media are often based on the assumption that the macroscopic constitutive properties can be taken as independent of the stress. This assumption makes possible explicit solutions for the stress field, but there is a growing realization \cite{tordesillas2009modeling,Bletal2015,matsushima2014universal,matsushima2017fundamental} that the local structure and stress self-organize cooperatively and preliminary observations were made on the stress-structure coupling \cite{Matsushima2021-li}. Such a coupling undermines the assumption of independence of local constitutive properties because any applied stress causes a correlated re-organisation of the local structure and suggests that the stress field equations {\it cannot be linear}.
While much effort focused on various stress-structure correlations, e.g., in terms of force chain structure \cite{SeeligWulff1946, oda1985stress, cates1998jamming, tordesillas2009modeling}, fabric tensors \cite{Rothenburg1989fabric, Bl04, Nguyen2015fabric}, and grain-scale structure \cite{tordesillas2010force, BaBl2002, Donev2007-lc}, the bulk effort focused on coarse-grained average descriptors and characterizing local effects of the stress-structure co-organization has been elusive. Here we quantify these effects in a range of planar systems, discuss the collapse of several related quantities onto a master curve, and provide a model that explains this collapse. \\ 
\indent In two dimensions, mechanically stable local structures can be described in terms of cells, which are the smallest voids surrounded by particles in contact~\cite{Ba96}, as illustrated in Fig.~\ref{fig:cellStructure}. The cell dynamics have proved useful to quantify kinetic processes \cite{kruyt2016micromechanical,wanjura2020structural}, and to study organization of structural characteristics and entropic effects \cite{tordesillas2010force,kuhn2014dense,matsushima2017fundamental,sun2020}.
By simulating a range of differently compressed planar systems, we study the self-organization of cells and the stress field. We define cell stresses and find that: 1. under simple scaling, the distributions of cell stress ratios of all the systems collapse onto one Weibull curve; 2. cells orient along the directions of the local principal stresses. We identify the effects of friction and cell order on these phenomena. Finally, we introduce a simple model that explains the stress-structure orientation correlations.
Our results establish and quantify the stress-structure coupled self-organization. They also show that a cell-based description is essential to the understanding of the structural evolution of granular materials. This study further suggests that current linear models of stress transmission in granular material may need to be modified.\\

\noindent{\bf Numerical simulations}:
The numerical simulations were carried out using the Discrete Element Method (DEM) \cite{pastrack1979discrete}. The initial state was generated by placing $N=50,000$ discs randomly within a double periodic domain. The discs' radii, $r$, were distributed uniformly between 0.667 and 1.333, with a mean $\bar{r}=1$. The initial packing fraction was set to below jamming, $\phi=0.7$, making the initial system dimensions $L_x=L_y\approx282.5\bar{r}$. 
The disc peripheries were assigned a friction coefficient $\mu$ and the system was slowly compressed by two different protocols: isotropic (ISO) and anisotropic (ANISO). We studied twelve systems, with each protocol applied with  $\mu=0.01,0.1,0.2,0.5,1.0$, and $10$. Systems were compressed by changing the periodic length in two directions in ISO and one direction in ANISO. During a compression step, disc center positions, $(x_i,y_i)$, were shifted using 
\begin{equation}
    \alpha_i(t+\Delta t) = \frac{L_\alpha(t+\Delta t)}{L_\alpha(t)} \alpha_i(t) \quad ; \quad \alpha_i = x_i, y_i \ .
\end{equation}
This ensured a uniform strain across the system.
The system's stress state is
\begin{equation}
    \Sigma_{\alpha\beta} = \frac{1}{L_x L_y}  \sum_{i<j} l^{ij}_\alpha f^{ij}_\beta  \ ,
\end{equation}
where $f^{ij}_\alpha$ is the $\alpha$-component of the contact force between discs $i$ and $j$ and $l^{ij}_\beta$ is the $\beta$-component of the position vector pointing from disc $j$'s center to that of disc $i$.
Compression continued until two conditions were met: (i) the system reached a jammed mechanically stable state, in which the kinetic energy fell below a very small magnitude $E_k<10^{-17} k_{n} \bar{r}^2$ per disc, where $k_n$ is the spring stiffness of our contact model ($=10^7$N/m in our simulations); 
(ii) the system stresses, $\Sigma_{xx}=\Sigma_{yy}$ in ISO and $\Sigma_{yy}$ in ANISO, reached a threshold value $\Sigma_c = k_n\bar{\delta} = 10^{2}$, with $\bar{\delta}$ the average overlap  between discs (limited to $10^{-5}\bar{r}$ to approximate disc rigidity).\\
\indent We then generated the coordination number histograms in each system, ignoring rattlers, and calculated the means, $\bar{z}$, and standard deviations, $\langle \delta^2 z\rangle$. These we summarize in Table \ref{tab:z}. The values of $\bar{z}$ ranged from $3.98$ for $\mu=0.01$ to $3.07$ for $\mu=10.0$ in both ISO and ANISO systems, showing that all the systems were very close to being marginally rigid.
For the ANISO processes, we also measured the stress ratios $\Sigma_{xx}/\Sigma_{yy}$ and found that these decreased from $0.948$ for $\mu=0.01$ to $0.815$ for $\mu=10.0$. \\
\begin{table}[!t]
    \centering
    \begin{tabular}{c|c c c c c c}
    \hline\hline
        $\bar{z}$ & $\mu=0.01$ & 0.1 & 0.2 & 0.5 & 1.0 & 10.0 \\
        \hline
        ISO   & 3.98 & 3.70 & 3.51 & 3.21 & 3.11 & 3.07 \\
        ANISO & 3.98 & 3.70 & 3.51 & 3.21 & 3.11 & 3.07 \\
    \hline\hline
    \end{tabular}
    
    \vspace*{0.3 cm}
    
    \begin{tabular}{c|c c c c c c}
    \hline\hline
        $\langle \delta^2 z\rangle$ & $\mu=0.01$ & 0.1 & 0.2 & 0.5 & 1.0 & 10.0 \\
        \hline
        ISO   & 0.85 & 0.83 & 0.83 & 0.84 & 0.84 & 0.82 \\
        ANISO & 0.86 & 0.82 & 0.83 & 0.84 & 0.83 & 0.82 \\
    \hline\hline
    \end{tabular}
    \caption{The systems' mean coordination numbers, $\bar{z}$, and standard deviations, $\langle \delta^2 z\rangle$.}
    \label{tab:z}
\end{table}

\begin{figure}[!h]
    \includegraphics[width=\linewidth]{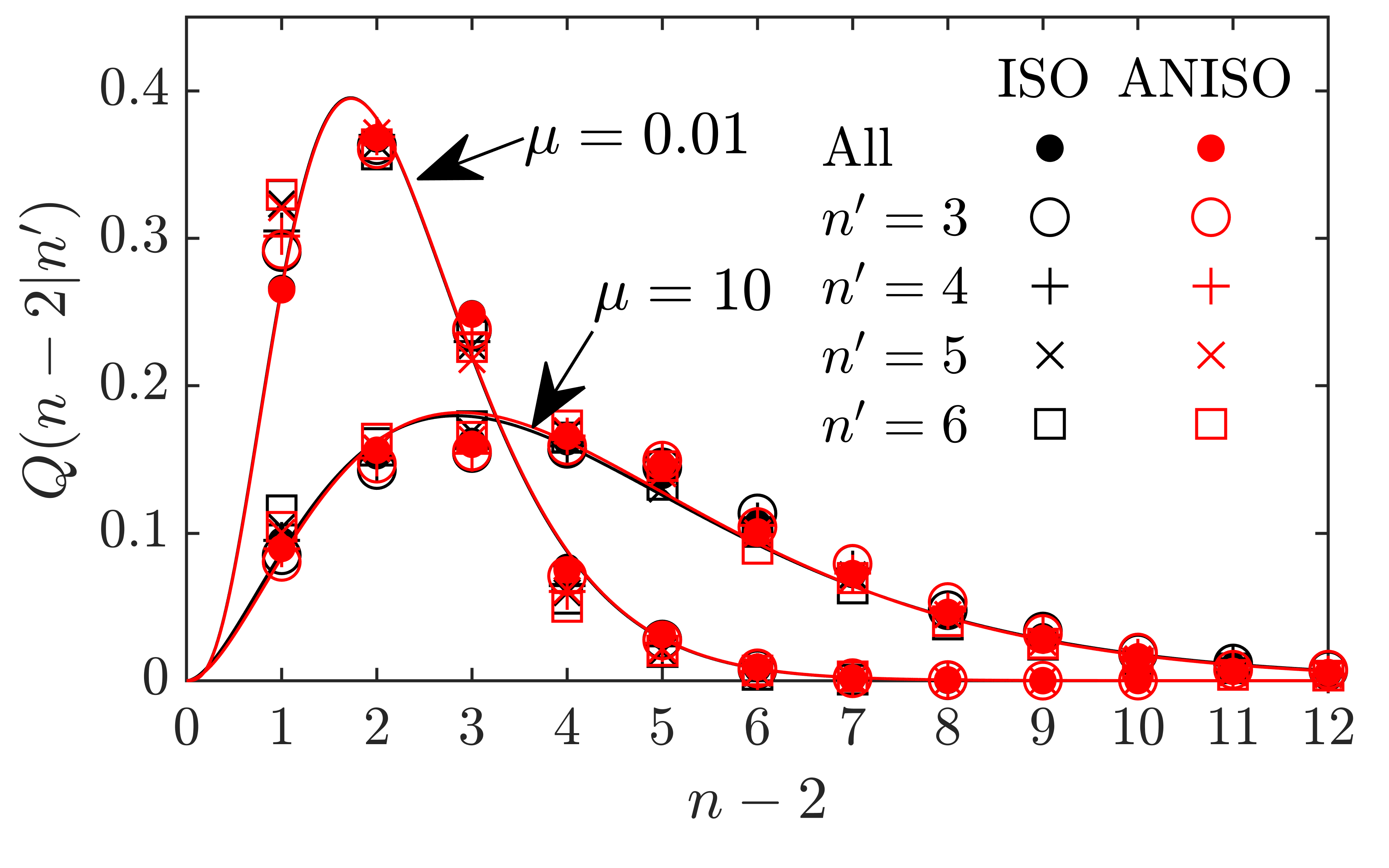}
    \caption{The conditional cell order PDFs, $Q(n|n')$, in all the simulated systems and for all cell orders, $n'$, collapse onto a single curve that depends only on $\mu$. Shown are typical collapsed curves for $\mu=0.01$ and $\mu=10.0$. The solid lines are $k-\Gamma$-distribution fits, eq. (\ref{kgamma}).}
    \label{fig:ccod}
\end{figure}

\begin{figure}[!h]
    \includegraphics[width=\linewidth]{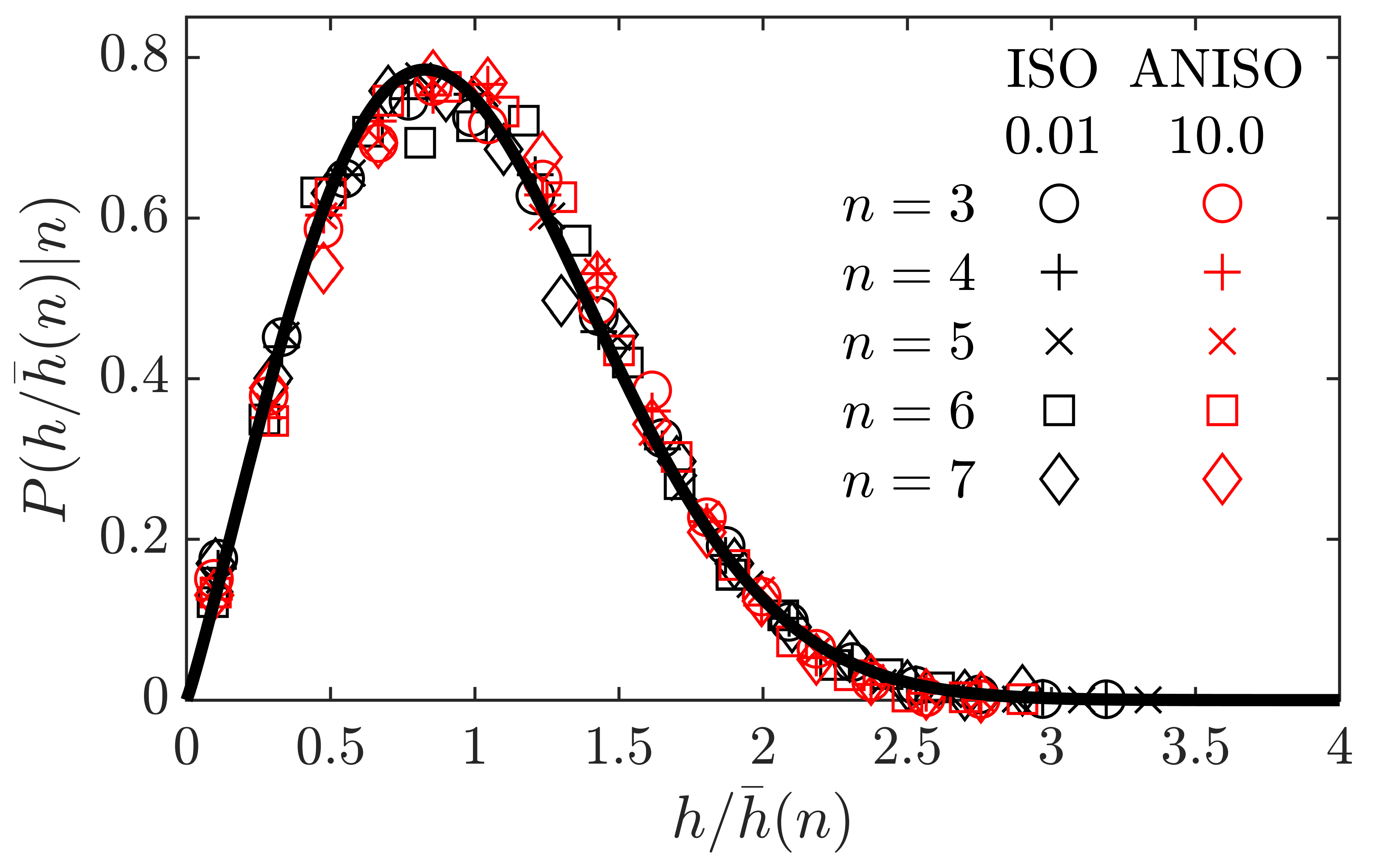}
    \caption{The conditional PDFs of $\hat{h}$ given $n$ collapse onto a single curve for all $n$ and $\mu$. The solid line represents the Weibull-distribution fit, eq. (\ref{eq:Weibull}).} 
    \label{fig:conditional pdf}
\end{figure}

\noindent{\bf Relations between cell order and stress}:
Our analysis of the generated packings is cell-based  \cite{BaBl2002} (see Fig.~\ref{fig:cellStructure}). 
Defining a cell center as the mean position of the contact points around it, we define a cell stress $\boldsymbol{\sigma}^c$ based on the stresses of the discs surrounding it. Let the stress of disc $g$ be
\begin{equation}
\boldsymbol{\sigma}^g = \frac{1}{A^g}  \sum_{g'} \boldsymbol{l}^{gg'} \otimes \boldsymbol{f}^{g'g}\ ,
\end{equation}
where $g'$ are the discs in contact with $g$, $\boldsymbol{l}^{gg'}$ the position vector pointing from the center of $g$ to the contact point with $g'$, $\boldsymbol{f}^{g'g}$ the contact force that $g'$ exerts on $g$, and $A^g$ is the area enclosed by the cell centers and contact points surrounding disc $g$, as illustrated in Fig.~\ref{fig:cellStructure}. 
The cell stress $\boldsymbol{\sigma}^c$ is
\begin{equation}
    \boldsymbol{\sigma}^c =  \sum_{g \in c} A^g  \boldsymbol{\sigma}^g \bigg{/}  \sum_{g \in c} A^g \ ,
\end{equation}
with $g \in c$ denoting the discs surrounding cell $c$. \\
\indent A cell order is the number of discs surrounding it~\cite{matsushima2014universal}. The static and dynamic cell order distribution (COD), $Q(n)$, were shown to have remarkable properties: maximum entropy \cite{sun2020} and detailed balance \cite{wanjura2020structural,sun2021experimental}, as well as providing a way to predict random close packing \cite{blumenfeld2021disorder}. 
Our measured CODs agree well with those reported in \cite{matsushima2014universal,matsushima2017fundamental} and, unlike in \cite{sun2020,blumenfeld2021disorder},  they are not exponential. The difference is because the COD is governed by a competition between mechanical stability and entropy maximisation \cite{sun2020}, with the latter dominant only at low external confining and driving forces -- a condition not satisfied in our compression processes. \\
As in \cite{wanjura2020structural,sun2021experimental}, we find that the orders of neighbor cells are uncorrelated and, furthermore, that this feature is independent of the process. i.e., the neighbor PDFs collapse to a single curve, which depends on the friction, $\mu$. This is illustrated in Fig.~\ref{fig:ccod} for the PDFs of $n$ around a cell of order $n'$, $Q(n|n')$ $\mu=0.01$ and $10.0$. The PDFs for the scaled variable $x\equiv(n-2)/\overline{(n-2)}$ are fitted well by a $k-\Gamma$ form
\begin{equation}
P(x)=\frac{k(\mu)^{k(\mu)} x^{k(\mu)-1}}{\Gamma[k(\mu)]}e^{-k(\mu)x}  \ ,  
\label{kgamma}
\end{equation}
with $k(\mu)$ decreasing monotonically from $4.2\pm 0.5$ at $\mu=0.01$ to $2.9\pm 0.3$ at $\mu=10.0$ (Fig. 5 in the supplemental material (SM) \cite{supmat22}). Typical collapsed curves and their fits are shown in Fig.~\ref{fig:ccod} for $\mu=0.01$ and $10.0$. \\
\indent Turning to cell stresses, the principal stresses are $\sigma^c_1$ and $\sigma^c_2<\sigma^c_1$, the mean and deviatoric cell stresses are, respectively, $p=(\sigma^c_1+\sigma^c_2)/2$ and $q=(\sigma^c_1-\sigma^c_2)/2$. 
We expect cell stability to be sensitive to the ratio $h=q/p$, which is supported by our model below. The PDFs of $h$ are found to be also independent of the packing process. With increasing $\mu$, they broaden and their mean, $\bar{h}$, increases (Fig. 6 in the SM \cite{supmat22}). \\
\indent To understand the PDF of $h$, we decompose it to its conditional probabilities in cells of order $n$ (henceforth $n$-cells), $P(h|n)$:
\begin{equation} 
    P(h) =  \sum_{n} Q(n)P(h|n) \ ,
    \label{eq:decomposition}
\end{equation}
where $Q(n)$ is the fraction of $n$-cells.
The mean at order $n$, $\bar{h}(n)=\sum_n h P(h|n)$, decreases with $n$ at roughly the same rate in all the systems, $\partial\bar{h}(n)/\partial n = -0.015\pm0.001$ (see the SM \cite{supmat22}).
This means that cell stability to deviatoric stress decreases with increasing cell order, consistent with the conclusions in \cite{matsushima2017fundamental} and \cite{sun2020}. 
We also observe that $\partial\bar{h}(n)/\partial\mu>0$, i.e., cell stability increases with friction. \\
\indent 
Defining $\hat{h}(n)=h(n)/\bar{h}(n)$ and rewriting (\ref{eq:decomposition}) as
\begin{equation}
P(h) = \sum_{n} Q(n)P(\hat{h}(n)|n)/\bar{h}(n) \ ,
\label{P2}
\end{equation}
we find that, remarkably, all the PDFs $P(\hat{h}(n)|n)$ collapse onto one curve, shown in Fig.~\ref{fig:conditional pdf}, independent of $n$, $\mu$, and compression protocol. This curve is fitted well by a single Weibull PDF:
\begin{equation}
    P\left(\hat{h}(n)|n\right) = C\frac{m}{\lambda} \left(\frac{\hat{h}(n)}{\lambda}\right)^{m-1} e^{-(\hat{h}(n)/\lambda)^m} \ ,
    \label{eq:Weibull}
\end{equation}
with  $\lambda=1.128\pm0.006$, and $m=2.09\pm0.015$ and $C=\left[1-e^{-(\hat{h}_{max}/\lambda)^m}\right]^{-1}$ a normalization constant. 
These collapses are indicative of an underlying self-organization and even possible universality \cite{matsushima2014universal,matsushima2017fundamental} and we study it in further detail below.
While there is a relation between the collapsed forms of $P\left(\hat{h}(n)|n\right)$ and the normalized distribution over all cells of $\hat{h}\equiv h/\left(Q_n\bar{h}(n)\right)$ (see SM~\cite{supmat22}), the two need not coincide. Nevertheless, we observe that the latter follows a similar Weibull form to very good accuracy. 

\begin{figure}[!h]
    \includegraphics[width=0.8\linewidth]{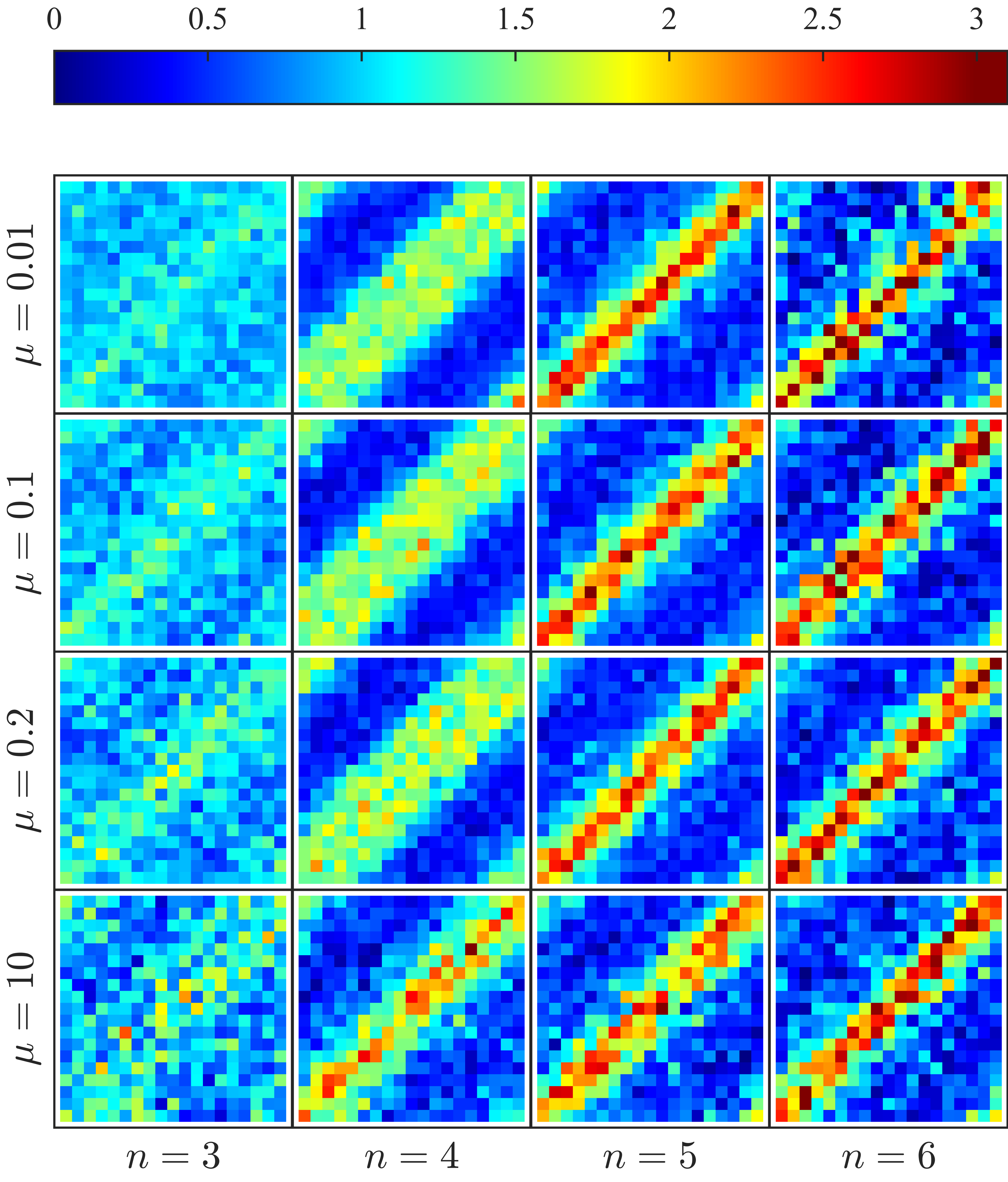}
    \caption{The conditional joint PDF of $\theta^\sigma$ and $\theta^c$ for cell orders $3\leq n\leq6$ in ISO systems exhibit a clear peak at $\theta^\sigma=\theta^c$.}
    \label{fig:ss correlation cond}
\end{figure}

\noindent{\bf Stress-structure self-organization}:\\
The cell configurations determines the local structure and it has been demonstrated that the co-organization is affected directly by the competition between maximizing configuration multiplicity and mechanical stability \cite{matsushima2017fundamental,sun2020}. To quantify the relation between stress and structural self-organizations, we first approximated each cell by its best fitted ellipse, using the method detailed in the SM \cite{supmat22}. We then studied the correlation between the cell ellipses' major axes, $\theta^c$, and their principal cell stress direction, $\theta^\sigma$. In ANISO, the orientations of $\theta^\sigma$ and $\theta^c$ follow the boundary stresses, which are not uniform (see Fig. 4 in the SM \cite{supmat22}) and, to eliminate this effect, we computed for each cell order the normalized joint PDFs $\frac{P(\theta^c ,\theta^\sigma |n)}{P(\theta^c|n)P(\theta^\sigma |n)}$, plotted in Fig.~\ref{fig:ss correlation cond} for ISO packings. The plots show a strong correlation between the cell orientations and stress principal in all the systems, except for $3$-cells, whose stability allows large orientational deviations from their stress principal axes and whose orientations are not well-characterized.\\
\begin{figure}[!h]
    \includegraphics[width=\linewidth]{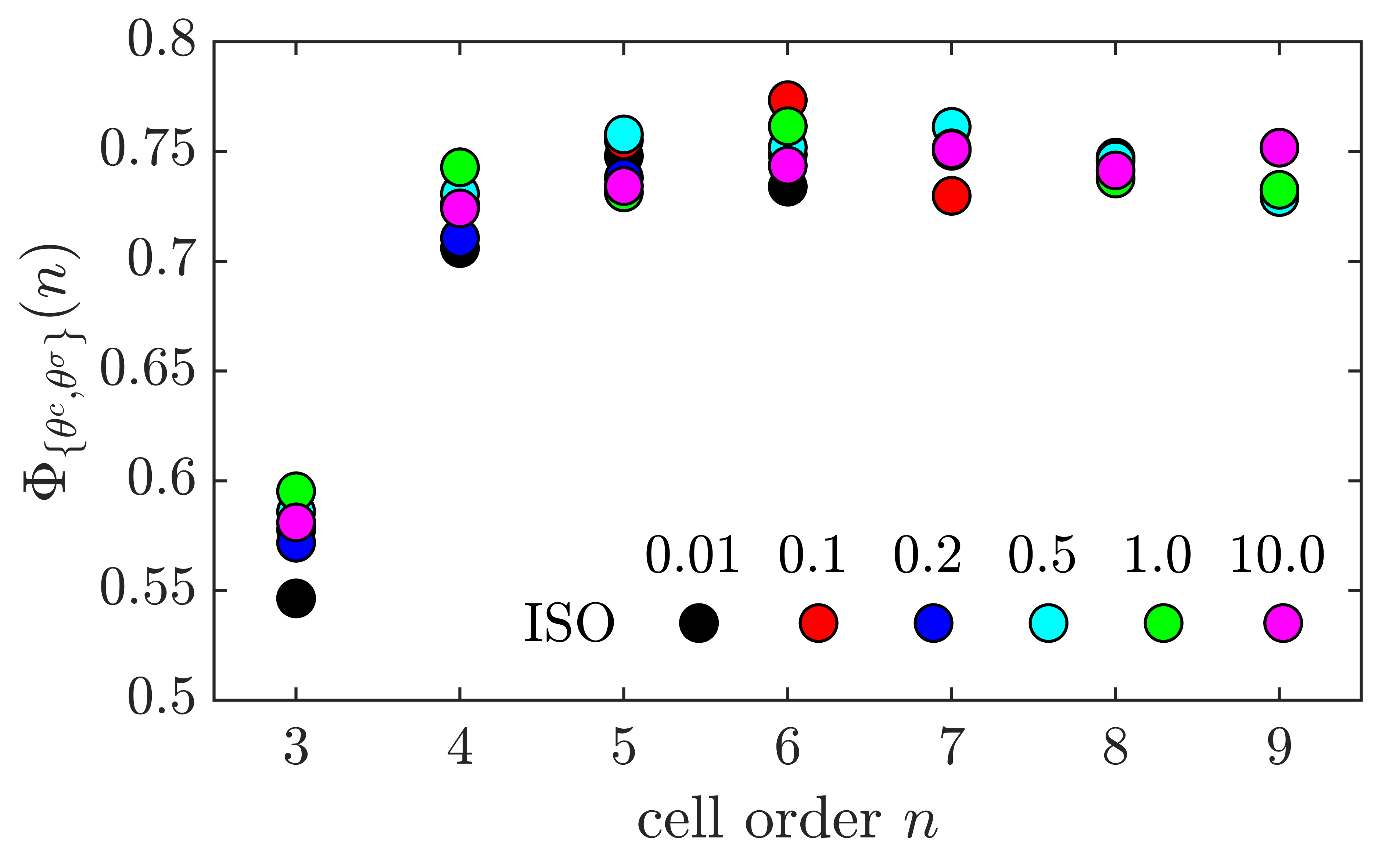}
    \caption{The correlation coefficients between $\theta^\sigma$ and $\theta^c$ given cell order $n$, $\Phi_{\{\theta^c,\theta^{\sigma}\}}(n)$, against $n$.}
    \label{fig:theta correlation coeff}
\end{figure}
\indent It is important to note that, while the peaks in these plots get sharper as $n$ increases, the plots are hardly affected by $\mu$. The plots for ANISO packings, shown in the SM \cite{supmat22}, exhibit identical behavior. Thus, cell stability and shapes appear independent of friction, suggesting that common wisdom concerning the effect of friction on these characteristics should be reconsidered.
We also show in the SM \cite{supmat22} the unconditional joint PDF, $\frac{P(\theta^c,\theta^\sigma)}{P(\theta^c)P(\theta^\sigma)}$, which behaves similarly.\\
\indent To quantify the stress-structure correlation, we computed the correlation coefficient between $\theta^c$ and $\theta^\sigma$ for each order, 
\begin{equation}
    \Phi_{\{\theta^c,\theta^{\sigma}\}}(n) =
    \frac{\sum_{i \in n} \delta\theta^c_i\delta\theta^\sigma_i}
    {\left[\sum_{i \in n} (\delta\theta^c_i)^2 \sum_{j \in n} (\delta\theta^{\sigma}_j)^2\right]^{1/2}},
\end{equation}
where $\delta\theta^c_i$ and $\delta\theta^\sigma_i$ are the differences from the respective mean angles at a given $n$.
Fig.~\ref{fig:theta correlation coeff} shows that the correlation is roughly the same for all systems, $\Phi=0.73\pm0.03$, except for $3$-disc cells, $\Phi=0.57\pm0.02$, which is lower for the reason discussed above.\\

\begin{figure}[!h]
    \centering
    \includegraphics[width=0.8\linewidth]{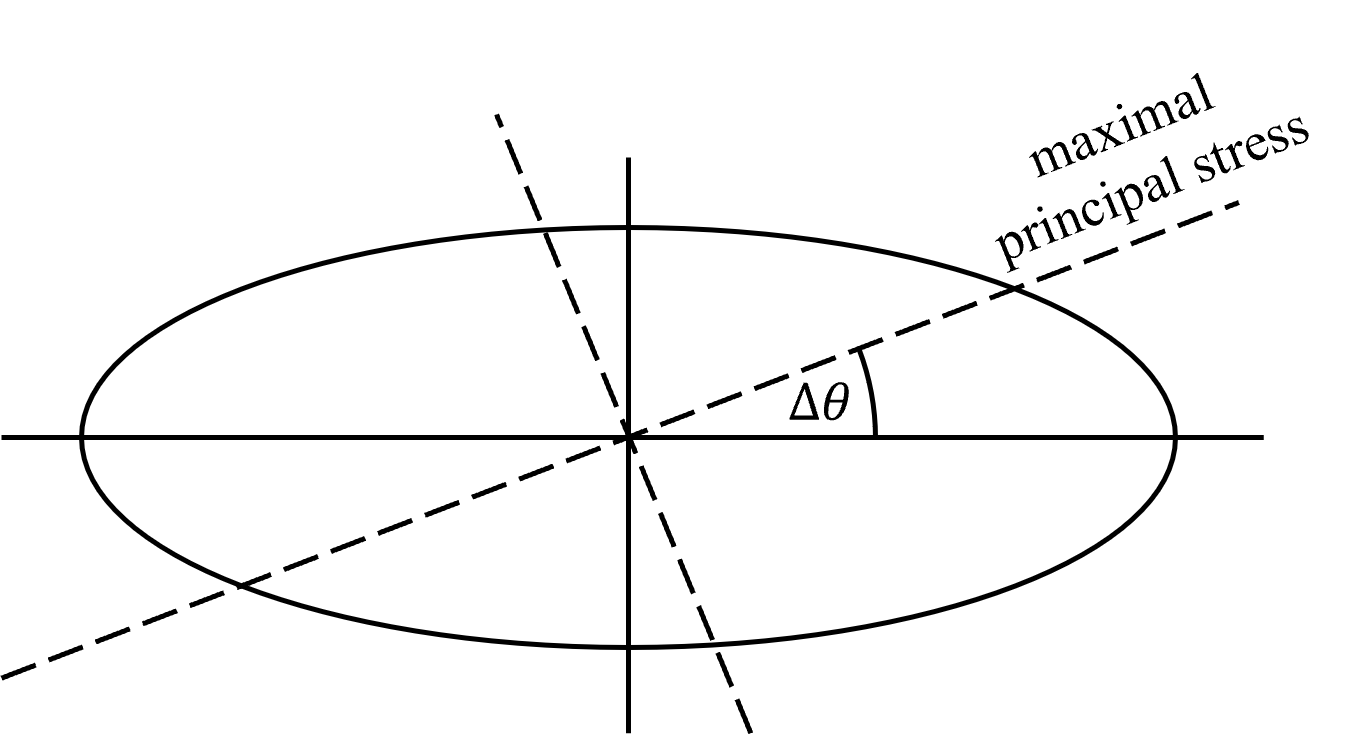}
    \caption{The local stress principal axes are tilted relative to the cell's equivalent ellipse by an angle of $\Delta\theta=\theta^\sigma-\theta^c$.}
    \label{CellEllipseStressModel}
\end{figure}

\noindent{\bf A model for the stress-structure organization}:
To understand the cooperative alignment of a cell with the local principal stress, consider its equivalent ellipse, whose major axis is at an angle $\Delta\theta=\theta^\sigma-\theta^c$ to the local principal static stress (Fig. \ref{CellEllipseStressModel}). In the reference frame aligned with the cell ellipse's major axis, the stress is
\begin{equation}
  \Sigma =  \begin{bmatrix}
        \sigma_1+\sigma_2\tan^2\Delta\theta & 
        2q\tan\Delta\theta \\
        2q\tan\Delta\theta & 
        \sigma_1\tan^2\Delta\theta+\sigma_2
    \end{bmatrix}\cos^2\Delta\theta \ .
\label{Stress1}
\end{equation}
\begin{figure}[!h]
    \centering
    \includegraphics[width=\linewidth]{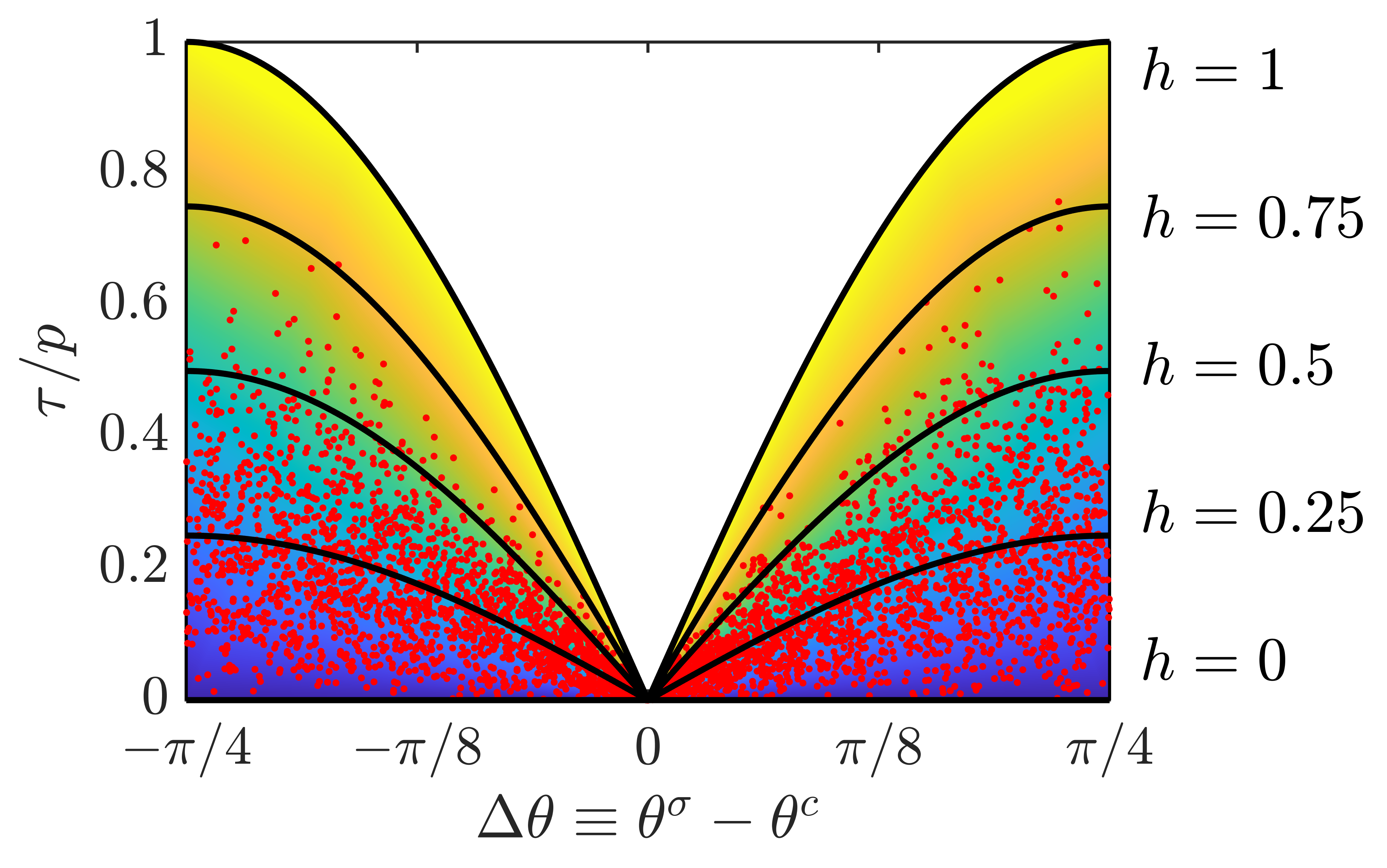}
    \caption{A scatter plot of cells in the $\Delta\theta-\frac{\tau}{p}$ parameter space. Each red point represents a cell and the background colour represents increasing value of $h$ from blue $h=0$ to yellow ($h=1$). The black lines are contours of constant $h$. The increasing cell density with low values $h$ and $\Delta\theta$ support the model that cell stress-structure alignment for stability is correlated with reduction in $h$.}
    \label{fig:CellTorqueModel}
\end{figure}
While the static stress is symmetric, the dynamics leading to the static state must involve force and torque imbalances to change cell configurations. Since the process is quasi-static, these imbalances perturb slightly the equilibrium stress. The more aligned are the cell major axis and the largest principal stress the more stable the cell is to compressive forces and it is plausible that imbalanced torque moments are more likely to reorient the cell when the (scaled) shear stress, $\tau/p=2q\sin{\Delta\theta}\cos{\Delta\theta}/p=h\sin{(2\Delta\theta)}$, is larger.
This is supported by the data shown in Fig. \ref{fig:CellTorqueModel}, in which each red point represents a cell with specific value of $h$ and ratio $\tau/p$. The data is clustered at low $\Delta\theta$ for any value of $h$ (black line contours), indicating the higher stability of cells with low shear stress. 

To understand the collapse in Fig~\ref{fig:ss correlation cond}, consider the steady-state distribution of $\hat{h}(n)$ ($0\leq\hat{h}\leq\hat{h}_{max}$) over the subset of $n$-cells. 
When an $n$-cell is created, the value of its $h’$ is not necessarily the most stable for its shape and, consequently, its configuration changes towards a stabler one, $h<h'$. Let the transition rate from $h'$ to $h$ be $f(h'\to h)$. Then
\begin{equation}
p_n(\hat{h}) = \int\limits_{\hat{h}}^{\hat{h}_{max}} p_n(\hat{h}') f(\hat{h}'\to \hat{h}) d\hat{h}' \ ,
\label{int}
\end{equation}
where we dropped the index $n$ for brevity. When the transition rates depend only on the target value, $f(\hat{h}'\to \hat{h})=g(\hat{h})$, this equation can be solved explicitly: 
\begin{equation}
p_n(\hat{h}) = Cg(\hat{h}) e^{-\int\limits^{\hat{h}} g(x)dx} \ ,
\label{DifEq}
\end{equation}
with $C$ a normalization constant. 
The collapse we observe is evidence that both $p_n$ and the rate, $g$, are independent of the order $n$. Moreover, the form of the distribution in (\ref{eq:Weibull}) provides a prediction of the explicit form of the transition rates: \\
\begin{equation}
g(\hat{h}) = {m \hat{h}^{m-1}}/{\lambda^m} \ ,
\label{Coll}
\end{equation}
a prediction that would be interesting to test in other granular dynamic processes. \\

\noindent{\bf Conclusions}:
To conclude, to study and quantify the local relation between stress and structure during the self-organization of granular systems, 
we simulated isotropic (ISO) and anisotropic (ANISO) compressions of several two-dimensional disc systems of interparticle friction coefficients in the range $0.01\leq\mu\leq10.0$. 
The local structure was characterized through the cell orders and orientations and cell stresses were defined. 
We have reported here the following. 
1. The conditional PDFs of the cell stress ratios at each order $n$, $h(n)=(\sigma_1^c-\sigma_2^c)/(\sigma_1^c+\sigma_2^c)$, collapse onto a single curve when scaled by their means, $\bar{h}(n)$. This master curve is fitted well by a Weibull PDF and it is independent of $n$, $\mu$, and packing protocol. 
2. A theoretical modelling of the collapse and the emergence of the Weibull distribution have been presented. The latter has been used to predict the transition rates of the stress ratio as cells rearrange. 
3. The mean stress ratio at cell order $n$, decreases with $n$ and increases with $\mu$, independently of the packing protocol.
4. The cell orientations and principal cell stress directions are strongly correlated and, excepting the isotropic $3$-cells, these correlations appear to be independent of $n$, $\mu$, and packing protocol. \\
\indent These results have several important implications. 
The independence of the collapsed conditional stress ratio PDFs of intergranular friction, a feature observed previously in collapses of the CODs~\cite{Matsushima2021-li}, is a fingerprint of self-organizations of the stress and the structure. The strong local correlation between the two further shows that the two self-organize cooperatively. Our results provide a quantitative measure of this phenomenon. 
Moreover, this feature could be universal in quasistatic granular processes and we suggest that our method be used to test for it in other systems. If universal, this feature would advance significantly understanding and modelling granular matter. \\
\indent 
The emergence of the Weibull distribution is particularly intriguing; to our knowledge, this is its first appearance in granular systems outside the contexts of failure, crushing, or survivability. \\
\indent Our cell-based approach to stress-structure relations can be readily extended to three dimensions. There, particle contacts around elementary voids make polyhedral cells, whose local structures can be quantified~\cite{BlEd06,Fretal09,Jo15}. Following methods analogous to those we used here, one can approximate the cell polyhedra by ellipsoids, define cell stresses, and investigate the correlations between the principal axes of cell ellipsoids and principal stresses. \\
\indent A significant implication of these results is that they call into question linear continuum stress theories for granular media because these assume that the medium's constitutive properties are independent of the stress. The correlations between the local structure means that local constitutive properties are also stress dependent. This, in turn, makes the stress equations nonlinear~\cite{Bletal2015}, suggesting that a new stress theory is necessary for granular media, which is based on local micromechanics.

\begin{acknowledgments}
\noindent RB acknowledges the financial support of the JSPS Invitational Fellowship ID=S22039 and the hospitality of Tsukuba University. 
TM acknowledges the financial support of JSPS KAKENHI Grants 20K20434A, 21H01422, and 21KK0071.
\end{acknowledgments}

\end{document}